\documentclass[11pt]{article}
\usepackage[latin1]{inputenc}
\usepackage{amsmath}
\usepackage{amsfonts}
\usepackage{amssymb}
\usepackage{color}
\usepackage{epsfig}
\newlength{\bredde}
\def\slash#1{\settowidth{\bredde}{$#1$}\ifmmode\,\raisebox{.15ex}{/}
\hspace*{-\bredde} #1\else$\,\raisebox{.15ex}{/}\hspace*{-\bredde} #1$\fi}
\newcommand{\be}{\begin{equation}}
\newcommand{\ee}{\end{equation}}
\newcommand{\bea}{\begin{eqnarray}}
\newcommand{\eea}{\end{eqnarray}}
\newcommand{\nn}{\nonumber}

\newcommand{\bs}{\begin{split}}
\newcommand{\es}{\end{split}}

\newcommand{\sect}[1]{\setcounter{equation}{0}\section{#1}}

\begin{document}
\topmargin -1.4cm
\oddsidemargin -0.8cm
\evensidemargin -0.8cm
\title{\Large\bf 
A Generalisation of Dyson's Integration Theorem for Determinants
}

\vspace{1.5cm}
\author{~\\{\sc G.~Akemann} and {\sc L.~Shifrin}
\\~\\
Department of Mathematical Sciences \& BURSt Research Centre\\
School of Information Systems, Computing and Mathematics\\
Brunel University West London\\ 
Uxbridge UB8 3PH, United Kingdom
}

\date{}
\maketitle
\vfill
\begin{abstract}
Dyson's integration theorem is widely used in the computation of eigenvalue
correlation functions in Random Matrix Theory. Here we focus on the variant 
of the theorem for determinants,  relevant for the  unitary ensembles with
Dyson index $\beta=2$. We derive a formula reducing the 
$(n-k)$-fold integral of an $n\times n$ determinant of a kernel of two sets 
of arbitrary functions to a determinant of size $k\times k$.
Our generalisation allows for sets of functions that are {\it not}
orthogonal or bi-orthogonal with respect to the integration measure. 
In the special case of orthogonal functions Dyson's theorem is recovered.


\end{abstract}
\vfill

\thispagestyle{empty}
\newpage

\renewcommand{\thefootnote}{\arabic{footnote}}
\setcounter{footnote}{0}


\sect{Motivation}\label{motiv}

Random Matrix Theory (RMT) has many applications in all areas of
Physics and beyond (see e.g. the introduction of \cite{Mehta}). For the class
of invariant RMT Dyson's integration 
theorem is at the heart of the method of orthogonal
polynomials when computing all 
eigenvalue correlation functions exactly, for finite
$n\times  n$ matrices. The resulting expressions are then amenable to the
large-$n$ limit, in which universal RMT predictions follow. 
In the following we restrict ourselves to the integration 
theorem for determinants. 
Before presenting our generalisation thereof we briefly recall 
how it reveals all eigenvalue correlations in the unitary ensembles.

We start by stating Dyson's integration theorem, as cited in \cite{Mehta}
(Theorem 5.1.4). 
Given $K(x,y)$ is a real valued function satisfying the following
self-contraction property: 
\bea
&&\int dy\  K(x,y)K(y,z) = K(x,z) \ , \nn\\
&& \int dy\ K(y,y) = c\ .
\label{dysonkernel}
\eea
Then it holds\footnote{The symmetry property $K(x,y)=K(y,x)$ stated in
  \cite{Mehta} is not necessary, as can be seen from the proof there.} 
\be
\int dx_1\ \det_{1\leq i,j\leq n}[K(x_i,x_j)] \ =\ (c-n+1)  
 \det_{2\leq i,j\leq n}[K(x_i,x_j)]\ ,
\label{dyson1}
\ee
thus reducing the size of the determinant by one through the integration. 
The theorem also holds for 
kernels of
orthogonal polynomials in the
complex plane or for bi-orthogonal polynomials.
A similar statement is true for quaternion valued kernels with the determinant 
replaced by a quaternion determinant (or Pfaffian). We refer to \cite{Mehta} 
for details as we will only consider the ordinary determinant case here. 
Iterating the integration theorem the following holds for an $(n-k)$-fold
integral: 
\be
\int \prod_{l=1}^{n-k}dx_l\ \det_{1\leq i,j\leq n}[K(x_i,x_j)] \ =\
(c-n+1)\ldots(c-k)
 \det_{n-k+1\leq i,j\leq n}[K(x_i,x_j)]\ .
\label{dysonk}
\ee
It is this form that we will generalise as it is most useful when computing 
correlation functions in RMT. We emphasise that on the right hand side (rhs)
the determinant has reduced to size $k\times k$ over the {\it same} kernel.

The application of eq. (\ref{dysonk}) to the unitary ensembles goes as follows.
Suppose we have a set of orthogonal polynomials
$p_k(x)=x^k+ O(x^{k-1})$ of order $k$ in monic normalisation 
satisfying
\be
\int dx\  w(x) p_k(x) p_j(x)\ =\ h_k \delta_{kj} \ .
\label{OPdef}
\ee
Here $w(x)$ is a positive weight function such that all moments exist.
From the polynomials we can construct orthonormal wave functions 
\be
\varphi_k(x)\ \equiv \ h_k^{-1/2} w(x)^{1/2} p_k(x)
\label{wavefunct}
\ee
and the following kernel 
\be
K_n(x,y)\ \equiv\ \sum_{j=0}^{n-1} \varphi_j(x) \varphi_j(y).
\label{OPkernel}
\ee
It satisfies Dyson's theorem above with $c=n$.
In the symmetry class of unitary invariant RMT, 
the partition function is given in terms of
the joint probability distribution (jpdf) of eigenvalues as 
\be
{\cal Z}_n= \int\prod_{i=1}^n dx_i\ w(x_i)\ \Delta_n(x)^2 \ .
\ee 
The Vandermonde determinant in the integrand, 
\be
\Delta_n(x)
= \det_{1\leq i,j\leq n}\left[x_i^{j-1}\right]
=  \det_{1\leq i,j\leq n}\left[p_{j-1}(x_i)\right]\ ,
\ee
can be replaced by a determinant over an arbitrary set of 
monic polynomials. If we choose the orthogonal ones we can rewrite 
the jpdf and thus the partition function after few manipulations as    
\be
{\cal Z}_n= \int\prod_{i=1}^n dx_i\ h_{i-1}
\det_{1\leq j,k\leq   n}[K_n(x_j,x_k)]. 
\label{Z}
\ee 
It immediately follows from 
Dyson's theorem that ${\cal Z}_n= n! \prod_{i=1}^n h_{i-1}$.
Moreover, following eq. (\ref{Z}) all $k$-point 
eigenvalue correlation functions given by $n-k$ integrations over the jpdf 
can be immediately read off:
\be
R_k(x_1,\ldots,x_k)\equiv \frac{1}{(n-k)!}\int\prod_{i=k+1}^n dx_i\ 
\det_{1\leq i,j\leq n}[K_n(x_i,x_j)]
\ =\ \det_{1\leq i,j\leq k}[K_n(x_i,x_j)]\ .
\ee
In the large-$n$ limit the size of the determinant on the rhs 
remains fixed, and the kernel can be easily evaluated 
using the Christoffel-Darboux identity for orthogonal polynomials on
$\mathbb{R}$.

Since in this example the choice of orthogonal polynomials was entirely at our
disposal, why 
should we choose polynomials that are not orthogonal with respect to the
weight function, or the integration range? 
The reason is that we are not always able to 
choose the polynomials to be orthogonal. 
One example where such a situation occurs is in the 
Schwinger model \cite{SV06}. A second example, being in a different symmetry
class, appears when
considering the Ginibre ensemble with 
real non-symmetric matrices \cite{AK}. 
Integrating out all real eigenvalues one arrives at the Pfaffian 
of the so-called $D$-kernel of the Gaussian Orthogonal Ensemble \cite{Mehta}, 
integrated over a non-Gaussian 
weight function in the complex plane. Consequently 
the self-contracting property eq. (\ref{dysonkernel}) is not satisfied then.  

For this reason we propose a generalisation of Dyson's theorem for
determinants without imposing any orthogonality condition,  
and we restrict ourselves to real integrals for simplicity. 
The generalisation 
to integrals over ${\mathbb C}$ is straightforward. 
A counterpart for integrating Pfaffians of a non self-contracting bilinear 
has been proved in  \cite{AK} in the
special case when all variables are integrated out.

\sect{Results}\label{results}
 
Let each $\{{\phi_j(p)}\}$ and $\{\psi_j(q)\}$, $j=1,\ldots,n$ 
be a set of linearly independent\footnote{We note that the functions 
$\phi_j(x)$ may or may not be linear combinations of the functions
  $\psi_j(x)$.},  integrable functions, such that all integrals $\int dx\
\phi_i(x) \psi_j(x)$ exist.  
For these two sets we define the following bilinear function:
\be
Q_n(x,y) \equiv\ \sum_{j=1}^{n} \phi_j(x) \psi_j(y)\ .
\label{kernelQ}
\ee
Then the following holds:\\

\underline{\sc Theorem 1:}
\be
\frac{1}{C}\int  \prod_{l=1}^{n-k}dx_l\ \det_{1\leq i,j\leq n}
\left[Q_n(p_i,q_j)\right] \ =\
(n-k)!   
\det_{n-k+1\leq i,j\leq n}[{\cal K}_n(p_i,q_j)]\ , \ \ k=0,\ldots,n\ ,
\label{Th1}
\ee
where we have set $p_i=q_i=x_i$, $i=1,\ldots, n-k$  for all integration
variables. 

The kernel ${\cal K}_n(p,q)$ on the rhs is given by 
\be
{\cal K}_n(p,q) \equiv\  \frac{1}{C}\sum_{a=1}^n \det
\left[
\begin{array}{lcr}
\int dx\ \phi_1(x) \psi_1(x) \cdots & 
\phi_1(p) \psi_a(q)
& \cdots \int dx\ \phi_1(x) \psi_n(x)\\
\hdots & \hdots & \hdots\\
\int dx\ \phi_n(x) \psi_1(x) \cdots & 
\phi_n(p) \psi_a(q)
& \cdots \int dx\ \phi_n(x) \psi_n(x)\\
\end{array}
\right],
\label{Kkernel}
\ee
where the sum runs over the $a$-th column replacing the 
integrated functions by unintegrated ones. The normalisation $C$ on the 
left hand side (lhs) is given by  
\be
C\ \equiv\ \det\left[\begin{array}{lr}
\int dx\ \phi_1(x) \psi_1(x) \cdots 
& \cdots \int dx\ \phi_1(x) \psi_n(x)\\
\hdots & \hdots\\
\int dx\ \phi_n(x) \psi_1(x) \cdots 
& \cdots \int dx\ \phi_n(x) \psi_n(x)\\
\end{array}
\right].
\label{norm}
\ee
Thus we have reduced an $(n-k)$-fold integral over an $n\times n$ determinant 
to a $k\times k$ determinant of a single kernel, consisting of  a sum of
$n\times n$ determinants containing only single integrals.

The rhs of our Theorem 1 can be interpreted as a generalised kernel having 
$2k$ 
variables. If we define 
\be
\label{kkerndef}
{\cal K}_n^{(k)}(p_1,\ldots,p_k;q_1,\ldots,q_k)\equiv \frac{1}{k!}
\det_{1\leq i,j\leq k}[{\cal K}_n(p_i,q_j)]\ , 
\ee
these satisfy the following generalised self-contraction property 
(see eq. (\ref{dysonkernel})):\\

\underline{\sc Theorem 2:}
\bea
\int dq_1\ldots dq_k\  {\cal K}_n^{(k)}(p_1,\ldots,p_k;q_1,\ldots,q_k) 
{\cal K}_n^{(k)}(q_1,\ldots,q_k;r_1,\ldots,r_k)&=&   
{\cal K}_n^{(k)}(p_1,\ldots,p_k;r_1,\ldots,r_k) \ , \nn\\
\int dq_1\ldots dq_k\  {\cal K}_n^{(k)}(q_1,\ldots,q_k;q_1,\ldots,q_k)&=& 
\bigg({n\atop k}\bigg) 
\ .
\label{self} 
\eea
In particular the kernel defined in eq. (\ref{Kkernel}), 
${\cal K}_n^{(k=1)}(p;q)\equiv{\cal K}_n(p;q)$, is self-contracting.

Let us make a few remarks. First, the bilinear $Q_n(x,y)$ of the set of
functions  
is in general {\it different} from the kernel on the rhs:
$Q(x,y)\neq  {\cal K}_n(x,y)$. In particular it is not
self contractive in general: 
$\int dy\  Q(x,y)Q(y,z) \neq Q(x,z)$.
In the case of orthogonal functions, 
$\int dx\,  \phi_k(x) \psi_j(x)\ =\ \delta_{kj}$, 
we obviously get back $Q_n(x,y)= {\cal K}_n(x,y)$. Then Dyson's theorem
applies, as in the example in the previous section.  

Special cases of Theorem 1 were previously known. For $k=0$ 
it goes back to C. Andr\'eief in 1883 as cited in \cite{TW}, after multiplying
with the normalisation $C$:
\be
\int  \prod_{l=1}^{n}dx_l\ \det_{1\leq i,j\leq n}
\left[\phi_j(x_i) \right]
\det_{1\leq i,j\leq n}\left[\psi_j(x_i)\right] \ =\
{n!}   
\det_{1\leq i,j\leq n}\left[\int dx\  \phi_i(x)\psi_j(x)\right]\ .
\label{andreief}
\ee 
The identity for $k=1$ was stated and used in \cite{SV06} but no explicit
proof was given.  
Furthermore let us point out that for $k=n$ there are no integrations, thus
equating the determinant of the bilinear function and of the kernel.

All statements above also hold when the normalisation eq. (\ref{norm})
accidentally vanishes, $C=0$, as will be indicated below. 
This cannot happen for Dyson's integration theorem.

\section{Proofs}

The proof of Theorem 1 will go in three steps, taking the known result for
$k=0$ for granted. 
In step i) we prove the Theorem for $k=1$, relating to the definition 
(\ref{Kkernel}). In step ii) we show that 
the kernel ${\cal K}_n(p,q)$
satisfies the self-contraction property eq. (\ref{self}), Theorem
2 for $k=1$.  
In the last step iii) we prove Theorem 1 for $k=n$ without
integrations. Because of  
the self-contraction property of ${\cal K}_n(p,q)$
we can then apply Dyson's theorem to the rhs to
show  all the remaining cases. 
Theorem 2 for $k\geq 2$ will then be shown in the second part.\\ 

\noindent
\underline{step i):}
To prove $k=1$,
in a first trivial step we can replace the determinant of the 
bilinear function as follows: 
\be
\det_{1\leq i,j\leq n}\left[
Q_n(p_i,q_j)
\right] \ =\  
\det_{1\leq i,j\leq n}\left[\sum_{a=1}^n \phi_a(p_i) \psi_a(q_j)\right] \ =\ 
\det_{1\leq a,i\leq n}\left[\phi_a(p_i)\right]
\det_{1\leq a,j\leq n}\left[\psi_a(q_j)\right].
\label{Qdecomp}
\ee
Inserting this into the lhs we can expand both determinants with respect 
to the last, unintegrated column:
\bea
\frac{1}{C}\int 
\prod_{l=1}^{n-1}dx_l\ 
\det_{1\leq a,i\leq n}\left[\phi_a(p_i)\right]
\det_{1\leq a,j\leq n}\left[\psi_a(q_j)\right] &=&
\frac{1}{C}\int \prod_{m=1}^{n-1}dx_m\ 
\left(\sum_{j=1}^n\phi_j(p_n)C^\phi_j\right)
\left(\sum_{l=1}^n\psi_l(q_n)C^\psi_l\right)
\nn\\
&=& \frac{(n-1)!}{C}
\sum_{l,j=1}^n \phi_j(p_n)\psi_l(q_n)C_{lj} \nn\\
&=& (n-1)!\ {\cal K}_n(p_n,q_n)\
,
\label{i}
\eea
where $p_i=q_i=x_i$ for $i=1,\ldots,n-1$.
We have introduced the minors
\be
C^\phi_j \ \equiv\ (-)^{n-1+j}\det_{i\neq j}[\phi_i(x_k)]\ \ \mbox{and}\ \ 
C^\psi_l \ \equiv\ (-)^{n-1+l}\det_{i\neq l}[\psi_i(x_k)]\ .
\ee
These contain $n-1$ functions each, and all variables $x_1,\ldots,x_{n-1}$ are
integrated.  
Thus for each product 
$C^\phi_j C^\psi_l$ we can apply the formula for $k=0$ 
by C. Andr\'eief eq. (\ref{andreief}), with the resulting minor
\be
C_{lj}\ \equiv\ (-)^{l+j}\det_{i\neq j;k\neq l}\left[\int dx\ \phi_i(x)
  \psi_k(x)\right]\ . 
\label{Cminor}
\ee
In the last step the sum in eq. (\ref{i}) 
can be precisely written as the sum over determinants in eq. (\ref{Kkernel}),
each expanded with respect to the $a$-th column.\\ 

\noindent
\underline{step ii):}
To derive the self-contraction property for ${\cal K}_n(p,q)$ we simply insert
 the definition eq.(\ref{Kkernel}), applying the short hand notation 
 $<i,k>\equiv\int dx\ \phi_i(x) \psi_k(x)$:
\bea
\int dq\  {\cal K}_n(p,q){\cal K}_n(q,r) &=& 
\frac{1}{C^2}\int dq
\left(
\sum_{a=1}^n \det
\left[
\begin{array}{lcr}
<1,1>\cdots & \phi_1(p) \psi_a(q)& \cdots <1,n>\\
\hdots & \hdots & \hdots\\
<n,1>\cdots & \phi_n(p) \psi_a(q) & \cdots <n,n>\\
\end{array}
\right]
\right)\nn\\
&&\ \ \ \ \ \ \ \ \ \times
\left(
\sum_{b=1}^n \det
\left[
\begin{array}{lcr}
<1,1>\cdots & \phi_1(q) \psi_b(r)& \cdots <1,n>\\
\hdots & \hdots & \hdots\\
<n,1>\cdots & \phi_n(q) \psi_b(r) & \cdots <n,n>\\
\end{array}
\right]
\right)\nn\\
&=& \frac{1}{C^2}\left(
\sum_{a=1}^n \det
\left[
\begin{array}{lcr}
<1,1>\cdots & \phi_1(p) \psi_{b=a}(r)& \cdots <1,n>\\
\hdots & \hdots & \hdots\\
<n,1>\cdots & \phi_n(p) \psi_{b=a}(r) & \cdots <n,n>\\
\end{array}
\right]
\right)\times C\nn\\
&=& {\cal K}_n(p,r)\ .
\label{ii}
\eea
Here we simply observe that in each product of determinants the common factors
$\psi_a(q)$ and $\psi_b(r)$ 
can be taken out of the columns $a$ and $b$ respectively, and can 
then be  multiplied into the columns of the 
other determinant. In a second step the integral 
$\int dq$ can now be taken inside the $b$-th column of the second determinant 
containing  $\phi_j(q) \psi_a(q)$, resulting into $<j,a>$. This leads to a
column already present and  
thus a vanishing determinant, unless we have $a=b$. The resulting 
normalisation cancels one power of $C$ to reproduce the kernel.\\

\noindent
\underline{step iii):}
We prove Theorem 1 
for $k=n$ when all integrations are absent. It is easily seen when expanding 
the kernel inside the determinant on the rhs, using the formula in the last
line of eq. (\ref{i}):  
\bea
\det_{1\leq i,j\leq n}[{\cal K}(p_i,q_j)]
&=&
\det_{1\leq i,j\leq n}\left[
\frac{1}{C}
\sum_{l,k=1}^n \phi_k(p_i)\psi_l(q_j)C_{lk}
\right]\nn\\
 &=&
\det_{1\leq i,k\leq n}\left[\phi_k(p_i)\right]
\det_{1\leq l,j\leq n}\left[\psi_l(p_j)\right]
\det_{1\leq l,k\leq n}\left[\frac{1}{C}C_{lk}\right] \nn\\
&=& 
\frac{1}{C}\ \det_{1\leq i,j\leq n}
\left[Q_n(p_i,q_j)\right] \ .
\label{iii}
\eea
In the first step we used that the determinant of the matrix product is the
product of the determinants.  
Furthermore, the minors are just the matrix elements of the inverse matrix,
$C_{lk}/C=(C^{-1})_{lk}$, and we used eq. (\ref{Qdecomp}).  
Because of ii) we can now apply Dyson's integration theorem to the
determinant of the self-contracting  
kernel ${\cal K}_n(p_i,q_j)$. Using its normalisation,  
$\int dq\ {\cal  K}_n(q,q)=n$ which  
can be trivially seen, we arrive at Theorem 1 for all $k=0,\ldots,n$.

Finally we deal with the case $C=0$. 
The Andr\'eief formula eq. (\ref{andreief}) still holds and 
vanishes identically. 
To make Theorem 1 nonsingular we multiply it with $C^k$.
Step i) proving $k=1$ remains true and in general nonzero,
using eq. (\ref{andreief}) for size $n-1$.
For $k\geq2$ the lhs of Theorem 1 times $C^k$ is zero. 
If the matrix under the determinant 
$C$ eq. (\ref{norm}) has rank $n-2$ or less, all 
determinants inside the kernel eq. (\ref{Kkernel}) vanish and make the rhs
identically zero. For rank $n-1$ it is easy to see that the 
kernel eq. (\ref{Kkernel}) factorises into two functions of $p$ 
and $q$, thus having a vanishing determinant on the rhs.\\

It remains to prove Theorem 2 for $k\geq 2$. The proof goes as follows. On the
rhs of  
(\ref{self}) we substitute each of the two kernels ${\cal{K}}_n^{(k)}$ as a
determinant  
of single kernels from their definition (\ref{kkerndef}). 
Using the standard representation
of a determinant, we get:
\be
\mbox{lhs} = \frac{1}{(k!)^2}\int dq_1...dq_k\sum_{\sigma,\sigma^\prime}
(-1)^{\sigma+\sigma^\prime}
\prod_{i=1}^k\bigg[{\cal{K}}_n(p_{\sigma(i)},q_i)
{\cal{K}}_n (q_i,r_{\sigma^\prime(i)})\bigg].
\ee
The fact that ${\cal{K}}_n(x,y)$ is self-contractive allows us to do all
the integrals over $q_i$, to obtain 
\bea
\mbox{lhs} = 
\frac{1}{(k!)^2}\sum_{\sigma,\sigma^\prime}(-1)^{\sigma+\sigma^\prime}
\prod_{i=1}^k{\cal{K}}_n \big(p_{\sigma(i)},r_{\sigma^\prime(i)}\big)
 = \frac{1}{k!}\det_{1\leq i,j\leq k}{\cal{K}}(p_i,r_j)
\equiv {\cal{K}}_n^{(k)}(p_1,...,p_k;r_1,...,r_k).
\eea
The normalisation of the generalised kernel ${\cal{K}}_n^{(k)}$ follows
directly from the normalisation of the single kernel ${\cal{K}}_n (p,q)$ 
which is $n$, and the repeated application of
Dyson's integration theorem to a $k\times k$ determinant integrated $k$
times. This way we get the binomial coefficient:
\be
\int dq_1\ldots dq_k\  {\cal K}_n^{(k)}(q_1,\ldots,q_k;q_1,\ldots,q_k)= 
\frac{n!}{(n-k)!\ k!}\ .
\ee 

The same arguments as at the end of the previous proof apply for $C=0$.

\section{Conclusions}

We have shown how to reduce any number of integrations over a determinant of 
a bilinear function of
non-orthogonal functions to a smaller determinant of a 
self-contracting kernel containing only
single integrals. This makes the large-$n$ limit feasible in such
a general setting at least in principle, 
given the single integrals can be evaluated. 
Due to the fermionic nature of the Vandermonde determinant 
other applications than the mentioned
Schwinger model should exist. Our result gives hope that an analogous 
Pfaffian integration theorem with some variables unintegrated also exists. 

After writing up this paper we learned from P. Forrester that the first part
of our result was derived independently by Rains \cite{R} in the context of
symmetrised increasing subsequences. His alternative proof is formulated 
in terms of the Pfaffian of an antisymmetric matrix kernel. In contrast, our
proof illuminates the close relation to Dyson's theorem.

\indent

\noindent
\underline{Acknowledgements}:
We would like to thank T. Guhr and E. Kanzieper for interesting comments, 
P. Forrester for pointing out references, and the Referee for his comment on
the degenerate case.
Part of this work was written up and presented during the workshop on 
``Random Matrix Theory: Recent Applications'' at the Niels 
Bohr Academy in May 2007, and we thank the organisers for the stimulating
atmosphere.  
Financial support by EPSRC grant EP/D031613/1 (G.A. and L.S.)
and European Community Network ENRAGE 
MRTN-CT-2004-005616 (G.A.) is gratefully acknowledged.


\end{document}